\documentclass[11pt]{article}
\usepackage{amsmath}
\begin{document}
\pagestyle{empty}

\begin{flushleft}
\Large
{SAGA-HE-118-97  
\hfill July 31, 1997}  \\
\end{flushleft}
 
\vspace{2.0cm}
 
\begin{center}
 
\LARGE{{\bf Antiquark Flavor Asymmetry }} \\
\vspace{0.2cm}

\LARGE{{\bf with New Accelerator Facilities  }} \\

\vspace{1.1cm}
 
\LARGE
{S. Kumano $^*$ }         \\
 
\vspace{0.4cm}
  
\LARGE
{Department of Physics}         \\
 
\vspace{0.1cm}
 
\LARGE
{Saga University}      \\
 
\vspace{0.1cm}

\LARGE
{Saga 840, Japan} \\

\vspace{1.5cm}
 
\LARGE
{Talk given at the International Workshop on} \\

\vspace{0.2cm}

{``Exciting Physics with New Accelerator Facilities", } \\

\vspace{0.4cm}

{Spring-8, Nishi-Harima, Japan, Mar.11-13, 1997} \\

{(talk on Mar.13, 1997)}  \\
 
\end{center}
 
\vspace{1.4cm}

\vfill
 
\noindent
{\rule{6.cm}{0.1mm}} \\
 
\vspace{-0.2cm}
\normalsize
\noindent
{* Email: kumanos@cc.saga-u.ac.jp. 
   Information on his research is available}  \\

\vspace{-0.6cm}
\noindent
{at http://www.cc.saga-u.ac.jp/saga-u/riko/physics/quantum1/structure.html.} \\

\vspace{+0.6cm}
\hfill
{\large to be published in proceedings by the World Scientific}

\vfill\eject
\setcounter{page}{1}
\pagestyle{plain}
\begin{center}
 
\Large
{Antiquark Flavor Asymmetry with New Accelerator Facilities} \\
 
\vspace{0.5cm}
 
{S. Kumano $^*$}             \\
 
{Department of Physics, Saga University}      \\

{Honjo-1, Saga 840, Japan} \\

\vspace{0.7cm}

\normalsize
Abstract
\end{center}
\vspace{-0.30cm}

Flavor asymmetry in light antiquark distributions is discussed.
In particular, recent progress on the $\bar u/\bar d$ asymmetry
is explained.
Then, we discuss possible future experimental studies.

\vspace{0.6cm}


\noindent
{\bf 1. Introduction}

\vspace{0.2cm}
Antiquark distributions had been assumed flavor symmetric
because quark-antiquark pairs are mainly created perturbatively.
However, it became possible to investigate
the flavor dependence in detail recently. 
The neutrino-induced dimuon events revealed that the $\bar s$
distribution is quite different from the light antiquark ones,
$\bar u$ and $\bar d$. Furthermore, we found that 
$\bar u$ and $\bar d$ are different in Gottfried-sum-rule
studies and in Drell-Yan processes.
Because $\bar u$ and $\bar d$ masses are very small compared
with the energy scales in deep inelastic processes,
the $\bar u/\bar d$ asymmetry is rather unexpected. 
However, it provides us a good opportunity for investigating
nonperturbative physics in the nucleon.
In this talk, we discuss the $\bar u$ difference from $\bar d$.

Although there were early indications on the 
light antiquark asymmetry in the SLAC,
EMC, and BCDMS data in the 1970's and 1980's, they are not conclusive
enough to state that the Gottfried sum rule is violated. It is either
because they could not estimate the small-$x$ contribution or because
the experimental errors are too large. The NMC paper in 1991 on
the Gottfried sum and on the flavor asymmetry
created a new topic on parton distributions.
Later, Drell-Yan results by the NA51
agreed with the NMC flavor asymmetry.
Furthermore, the preliminary Fermilab-E866 
and HERMES semi-inclusive data tend to agree with the NMC flavor
asymmetry. 
On the other hand, various theoretical ideas were proposed. 
The most promising one is a meson-cloud picture; however, 
there could be also Pauli-blocking-type effects.

The following $\bar u/\bar d$ asymmetry discussions are based
on the summary paper in Ref. [1]. 
We discuss present theoretical and experimental results first,
then future possibilities at new accelerator facilities are discussed.
In this paper, we focus much on experimental aspects. 
If the reader is interested in learning more about the experimental
details or about theoretical studies, the author suggests one
to read the summary paper [1].

\vspace{0.6cm}

\noindent
{\bf 2. Present situation}

\vspace{0.2cm}

The light antiquark distribution $\bar u-\bar d$
could be studied in various processes. 
So far, two major experimental methods have been used.
One is the Gottfried sum in electron or muon deep inelastic 
scattering, and the another is the Drell-Yan process.
After the Gottfried proposal in 1967, the sum rule has a long
history of experimental investigations. However, the situation
was not clear enough before the NMC data in 1991. As it is common
in most sum rules, the small $x$ information is crucial for finding
an accurate sum.
The sum is related to the $F_2$ structure functions for the proton
and neutron. It is written in a parton model as
\begin{equation}
I_G=\int_0^1 \frac{dx}{x} \, [ F_2^p(x,Q^2) - F_2^n(x,Q^2) ]
 = \frac{1}{3} + \frac{2}{3} 
                 \int_0^1 dx [ \bar u(x,Q^2) - \bar d(x,Q^2) ]
\ \ .
\label{eqn:GINT}
\end{equation}
If the $\bar u$ distribution is equal to $\bar d$, 
the second term vanishes and the above equation becomes
the Gottfried sum $I_G=1/3$.
According to the NMC reanalysis in 1994, the sum is
\begin{equation}
I_G = 0.235 \pm 0.026
\ \ .
\end{equation}
This is significantly smaller than
the predicted value, and a natural way of interpreting
the deficit is to assume a $\bar d$ excess over $\bar u$
according to Eq. (\ref{eqn:GINT}).

An independent experimental result was reported by the NA51 group
in 1994 by using Drell-Yan data:
\begin{equation}
\bar u/\bar d = 0.51 \pm 0.04 (stat.) \pm 0.05 (syst.)
\ \ \ 
{\rm at}\ x=0.18
\ \ . 
\end{equation}
Unfortunately, the data point is given at only one
$x$ point, much detailed studies are needed. Currently,
the same experiment is going on at Fermilab by the
E866. Because the previous Fermilab experiments with
nuclear targets did not find significant asymmetry, accurate
results are awaited.

There are not so many  theoretical publications in 1970's and 1980's.
There are some discussions based on a Pauli-blocking idea and 
on a diquark picture. It should be noted that perturbative QCD cannot
explain the large deficit in the Gottfried sum.
The next-to-leading (NLO) correction is calculated as
\begin{align}
I_G &= \frac{1}{3} \left [ 1 + 
      \frac{3(C_F C_A/2 - C_F^2)}{2(33-2n_f)} \, 
       ( 13+8 \zeta (3)-2 \pi^2 ) \,   
      \frac{\alpha_s (Q^2)}{\pi} 
     \right ]
\nonumber \\
&= \frac{1}{3} \left [ 1 +    
\begin{pmatrix} 0.03552 \, (n_f=3) \\
                0.03836 \, (n_f=4) 
\end{pmatrix}
              \frac{\alpha_s (Q^2)}{\pi} \right ]
\ \ \ .
\label{eqn:ignlo}
\end{align}
The NLO contribution is merely 0.3\% at $Q^2$=4 GeV$^2$, and it is
too small to explain the 30\% deficit.
Therefore, the interpretation should originate from nonperturbative
physics. Although there is an estimate of the asymmetry
in lattice QCD, a complete calculation is still too far away.
Therefore, we should conjecture a possible model for explaining
the experimental results. Various ideas are proposed. 
After several years since the NMC discovery, 
the situation seemed to settle down.
The meson-cloud model is the most promising in the sense that
the major part of the NMC and NA51 could be explained.
Because of the difference between
$\bar u$ and $\bar d$ in virtual meson
clouds in the nucleon, we have the flavor asymmetry.
For example, the proton decays into $\pi^+ n$ or $\pi^0 p$.
Because the $\pi^+$ has a valence $\bar d$ quark,
these processes produce an excess of $\bar d$
over $\bar u$ in the proton.
The other idea is based on the Pauli-exclusion principle.
According to this model, $u\bar u$ pair creations are more suppressed
than $d\bar d$ creations because of the valence $u$ quark excess
over valence $d$ in the proton. 
However, it seems that the effects are not large enough to explain
the whole asymmetry. There are several proposed models; however,
these are the major explanations. Because theoretical models
are not discussed further in this paper, the interested reader may
read Ref. [1].

\vspace{0.6cm}
\noindent
{\bf 3. Future $\bf \bar u/\bar d$ asymmetry studies}

\vspace{0.4cm}
\noindent
{\bf 3.1 Drell-Yan process}
\vspace{0.2cm}

There are data which have been already taken in connection with
the flavor asymmetry; however, they are not still published.
First, accurate Drell-Yan data should be reported by the Fermilab
E866. The another is the HERMES semi-inclusive data,
which are discussed in the next subsection.

The Drell-Yan is a lepton-pair production process in hadron-hadron
collisions $A+B\rightarrow \ell^+\ell^- X$.
In the parton model, it is described by quark-antiquark
annihilation processes $q+\bar q\rightarrow \ell^+\ell^-$.
The cross section is given by\begin{equation}
s \frac{d\sigma}{d\sqrt{\tau}dy} =
\frac{8\pi\alpha^2}{9\sqrt{\tau}} \sum_i \, e_i^2 \,
[ \, q_i^A(x_1,Q^2) \, \bar q_i^B(x_2,Q^2)
  + \bar q_i^A(x_1,Q^2) \, q_i^B(x_2,Q^2) \, ]
\ ,
\label{eqn:DYCROSS}
\end{equation}
where $Q^2$ is the dimuon mass squared: $Q^2=m_{\mu\mu}^2$, 
and $\tau$ is given by $\tau=m_{\mu\mu}^2/s=x_1 x_2$.
According to the above equation, the process
can be used for measuring the antiquark distributions if 
the quark distributions in another hadron are known.
For finding the flavor asymmetry $\bar u-\bar d$,
the difference between p-p and p-n (practically p-d)
Drell-Yan cross sections is useful. 
Considering the rapidity point $y$=0
and retaining only the valence-sea annihilation terms, we have
\begin{align}
\sigma^{pp} &= \frac{8\pi\alpha^2}{9\sqrt\tau} \,
                 \left [ \, \frac{8}{9} u_v (x) \bar u (x)
                        +\frac{2}{9} d_v (x) \bar d (x)
                 \, \right ]
\ , \nonumber \\
\sigma^{pn} &= \frac{8\pi\alpha^2}{9\sqrt\tau} \,
       \left [ \, \frac{4}{9} \left \{ u_v (x) \bar d (x) 
                                   +d_v (x) \bar u (x) \right \}
             + \frac{1}{9} \left \{ d_v (x) \bar u (x) 
                                   +u_v (x) \bar d (x) \right \} 
        \, \right ]
\ ,
\end{align}
for the proton-proton and proton-neutron cross sections.
All the above distributions are at $x=\sqrt\tau$
because of $y=0$.
From these equations, the p-n asymmetry becomes
\begin{align}
A_{DY} &= \frac{\sigma^{pp}-\sigma^{pn}}{\sigma^{pp}+\sigma^{pn}}
              \nonumber \\
       &= \frac{[4u_v(x)-d_v(x)][\bar u(x)-\bar d(x)]
                +[u_v(x)-d_v(x)][4\bar u(x)-\bar d(x)]}
                {[4u_v(x)+d_v(x)][\bar u(x)+\bar d(x)]
                +[u_v(x)+d_v(x)][4\bar u(x)+\bar d(x)]}
\ \text{at $y=0$}
\ .
\end{align}
This quantity is very sensitive to the $\bar u-\bar d$ distribution.
However, antiquark-quark annihilation processes also contribute to
the above equation at the rapidity point $y$=0. 
In this sense, it is better to take large $x_F$ ($\equiv x_1-x_2$)
data so that the antiquarks in the projectile do not affect 
the asymmetry:
\begin{equation}
A_{DY} =  \frac {[4u(x_1)-d(x_1)][\bar u(x_2)-\bar d(x_2)]}
                {[4u(x_1)+d(x_1)][\bar u(x_2)+\bar d(x_2)]}
\ \ \ \text{at large $x_F$}
\ \ \ .
\label{eqn:ADYKL}
\end{equation}
It is obvious from these expressions, $A_{DY}$ can be used to measure
the $\bar u-\bar d$ distribution.
Currently, experimental analysis is in progress by the Fermilab-E866.
Its preliminary results indicate the $\bar u$ and $\bar d$
distributions which are consistent with 
the NMC type flavor asymmetry.
Although we should wait for the final report, the data seem to
confirm the NMC finding.

Because there is no neutron target, the deuteron is used
for obtaining the asymmetry $A_{DY}$. There is also a possibility
to investigate this quantity at RHIC if the deuteron is accelerated.
Furthermore, the flavor asymmetry in polarized parton distributions
should be an interesting topic to be investigated in polarized
reactions.

\vspace{0.4cm}
\noindent
{\bf 3.2 Charged-hadron production}
\vspace{0.2cm}

Charged-hadron production processes could have information
on the flavor asymmetry. The EMC semi-inclusive data were
used for studying a relation to the Gottfried sum. However,
the errors are too large to find a significant flavor
asymmetry effect. Although it is not published yet, the 
HERMES collaboration used its charged-pion data
to find the asymmetry. 
The charged-hadron cross section is proportional to the factor
\begin{align}
N^{Nh^\pm} &\equiv \sum_i e_i^2 \, f_i(x) \, D_i^{h^\pm}(z)    
   \nonumber \\
       &= \frac{4}{9} \, u \, D_u^\pm 
        + \frac{4}{9} \, \bar u \, D_{\bar u}^\pm
        + \frac{1}{9} \, d \, D_d^\pm 
        + \frac{1}{9} \, \bar d \, D_{\bar d}^\pm 
        + \frac{1}{9} \, s \, D_s^\pm 
        + \frac{1}{9} \, \bar s \, D_{\bar s}^\pm 
\ \ \ ,
\end{align}
where $f_i(x)$ is the quark distribution with flavor $i$
and momentum fraction $x$,
and $D_i^h(z)$ is the $i$-quark to $h$-hadron
fragmentation function with $z=E_h/\nu$.
Assuming the isospin symmetry in the parton distributions, 
we consider a combination of proton and neutron cross sections:
\begin{align}
R(x,z) &= \frac{(N^{p+}-N^{n+})+(N^{p-}-N^{n-})}
                {(N^{p+}-N^{n+})-(N^{p-}-N^{n-})} \nonumber \\
       &= \frac{u(x)-d(x)+\bar u(x)-\bar d(x)}
                {u(x)-d(x)-\bar u(x)+\bar d(x)} \cdot
           \frac{4 \, D_u^+(z)+4 \, D_{\bar u}^+(z)
                 -D_d^+(z)- D_{\bar d}^+(z)}
                {4 \, D_u^+(z)-4 \, D_{\bar u}^+(z)
                 -D_d^+(z)+ D_{\bar d}^+(z)}
\ .
\end{align}
The $\bar u-\bar d$ distribution could be extracted
if we have information on the fragmentation functions.
The recent HERMES preliminary data seem to be accurate enough
to find the $\bar u/\bar d$ asymmetry.
The following $\pi^+$ and $\pi^-$ production ratio is related to the 
function $R(x,z)$ for the pion by
\begin{equation}
r(x,z) = \frac{N^{p\pi^-} - N^{n\pi^-}}{N^{p\pi^+} - N^{n\pi^+}}
       = \frac{R_\pi (x,z) -1}{R_\pi (x,z)+1}
\ \ \ .
\end{equation}
The obtained data of $r(x,z)$ in the range $0.1 < x < 0.3$
agree well with the NMC flavor asymmetry, and they are 
significantly different from the symmetric expectation.
The HERMES results will be submitted for publication
in the near future.

\vspace{0.4cm}
\noindent
{\bf 3.3 W charge asymmetry}
\vspace{0.2cm}

W production processes should be also useful in finding
the flavor asymmetry. In particular, the W-charge-asymmetry
studies at RHIC should be important not only in unpolarized
distributions but also in polarized ones.
In the following, we discuss the unpolarized case.
We write $W^+$ production cross section in the $p+p$ reaction
in terms of the parton distributions:
\begin{align}
\frac{d \sigma_{p+p\rightarrow W^+}}{dx_F}  
                 =  K \, \frac{\sqrt 2 \pi}{3} \, G_F
\left(    \frac{x_1 x_2}{x_1 + x_2}   \right)
& \left\{  \, \cos^2 \theta_c  \,
[u(x_1) \bar d(x_2) + \bar d(x_1) u(x_2)]  \right. \nonumber \\
& \left. + \sin^2 \theta_c  \,
[u(x_1) \bar s(x_2) + \bar s(x_1) u(x_2)] \, \right\}
\ \ \ .
\end{align}
The dominant processes of producing $W^+$ are
$u(x_1)+\bar d(x_2)\rightarrow W^+$ and 
$u(x_2)+\bar d(x_1)\rightarrow W^+$; however,
the first one becomes much larger than the second at large $x_F$.
Therefore, the cross section is
sensitive to the $\bar d$ distribution at large $x_F$. 
On the other hand,
the cross section for the $W^-$ production is given 
in the same way:
\begin{align}
\frac{d \sigma_{p+p\rightarrow W^-}}{dx_F} 
                 = K \, \frac{\sqrt 2 \pi}{3} \, G_F
\left( \frac{x_1 x_2}{x_1 + x_2}\right)
& \left\{ \, \cos^2 \theta_c \,
[\bar u(x_1) d(x_2) +  d(x_1) \bar u(x_2)] \right. \nonumber \\
& \left. + \sin^2 \theta_c \,
[\bar u(x_1) s(x_2) + s(x_1) \bar u(x_2)] \, \right\}
\ \ \ .
\end{align}
At large $x_F$, it is sensitive to the $\bar u$ distribution
instead of the $\bar d$ in the $W^+$ case.
This difference makes it possible to find the distribution
$\bar u-\bar d$.

There are existing data on the W charge asymmetry by the CDF;
however, they are taken in the $p+\bar p$ reaction.
If we follow the similar equations to the above ones
for the $p+\bar p$ reaction, we find that the $p+\bar p$
cross sections are not sensitive to $\bar u/\bar d$.
Therefore, it is very important to have the asymmetry data
in the $p+p$ reaction. The RHIC is certainly a possible 
candidate to measure this quantity.

\vspace{0.4cm}
\noindent
{\bf 3.4 Deuteron acceleration at HERA}
\vspace{0.2cm}

As a future project of HERA, acceleration of some nuclei
is considered although it is not clear whether such a project
is approved. If the electron-deuteron collider is realized at HERA,
we should be able to test the NMC finding directly. The smallest
$x$ point in the NMC experiment is $x=0.004$. Although they estimated
the smaller $x$ contribution by extrapolation, it is not very obvious
whether the small $x$ contribution is small.
In order to test it directly, we need a high-energy accelerator.
If it could be done at HERA, they should be able to get the data
in the region $x=10^{-4}-10^{-5}$. These data are essential for
finding the small $x$ contribution to the Gottfried sum.

\vspace{0.00cm}

\begin{center}
{\bf Acknowledgments} \\
\end{center}
\vspace{-0.17cm}

S. K. thanks RCNP for its financial support 
for participating in this conference.
He thanks the theory group at SLAC where 
this manuscript is written.

\vspace{0.2cm}

\noindent
{* Email: kumanos@cc.saga-u.ac.jp. 
   Information on his research is available}  \\

\vspace{-0.55cm}
\noindent
{\ \ \, at http://www.cc.saga-u.ac.jp/saga-u/riko/physics/quantum1/structure.html.} \\

\vspace{-0.30cm}

\begin{center}
{\bf Reference} \\
\end{center}
 
\vspace{-0.20cm}

\begin{description}{\leftmargin 0.0cm}

\vspace{-0.20cm}
\item{[1]}
S. Kumano, preprint SAGA-HE-97-97 (hep-ph/9702367).
Papers on the $\bar u/\bar d$ asymmetry are found
in the reference section of this preprint.


\end{description}

\end{document}